\documentclass{article}
\usepackage{amsmath,graphicx,mlspconf,xcolor,bm}
\usepackage{amssymb}
\usepackage{subcaption}

\toappear{2024 IEEE International Workshop on Machine Learning for Signal Processing, Sept.\ 22--25, 2024, London, UK}

\title{TASK-ORIENTED COMMUNICATION FOR VEHICLE-TO-INFRASTRUCTURE COOPERATIVE PERCEPTION}
\name{Jiawei Shao, Teng Li, Jun Zhang}
\address{
The Hong Kong University of Science and Technology, Hong Kong \\
Email: \{jiawei.shao, tliby\}@connect.ust.hk, eejzhang@ust.hk}

\begin{document}

    \maketitle

\begin{abstract}

Vehicle-to-infrastructure (V2I) cooperative perception plays a crucial role in autonomous driving scenarios. 
Despite its potential to improve perception accuracy and robustness, the large amount of raw sensor data inevitably results in high communication overhead. 
To mitigate this issue, we propose TOCOM-V2I, a task-oriented communication framework for V2I cooperative perception, which reduces bandwidth consumption by transmitting only task-relevant information, instead of the raw data stream, for perceiving the surrounding environment.
Our contributions are threefold. 
First, we propose a spatial-aware feature selection module to filter out irrelevant information based on spatial relationships and perceptual prior.
Second, we introduce a hierarchical entropy model to exploit redundancy within the features for efficient compression and transmission.
Finally, we utilize a scaled dot-product attention architecture to fuse vehicle-side and infrastructure-side features to improve perception performance.
Experimental results demonstrate the effectiveness of TOCOM-V2I.

\end{abstract}
\begin{keywords}
Task-oriented communication, cooperative perception, entropy model
\end{keywords}

\section{Introduction}

Accurate environmental perception is paramount in the development of autonomous vehicle technology. 
Currently, self-driving vehicles heavily rely on onboard sensors, such as LiDAR, Radar, and cameras, to obtain a precise 3D representation of their surroundings \cite{paden2016survey}.
However, due to the limited sensing range and field of view of on-board sensors, the vehicle perception system faces challenges such as occlusions, blind spots, and difficulty in detecting distant or small objects \cite{hu2022where2comm}. 
These limitations result in incomplete perception data, potentially compromising safety and decision-making capabilities in complex driving scenarios.
In response to these issues, recent studies \cite{liu2020when2com,liu2020who2com} have been focusing on the integration of data from both the vehicle and the roadside infrastructure to achieve V2I cooperative perception.
Such a cooperative approach extends the sensing capabilities of individual vehicles by incorporating perceptual information from infrastructure-mounted sensors.
This allows for a more comprehensive view of the environment, enabling vehicles to detect objects beyond their field of view.

\begin{figure}
    \centering
    \includegraphics[width = 0.99\linewidth]{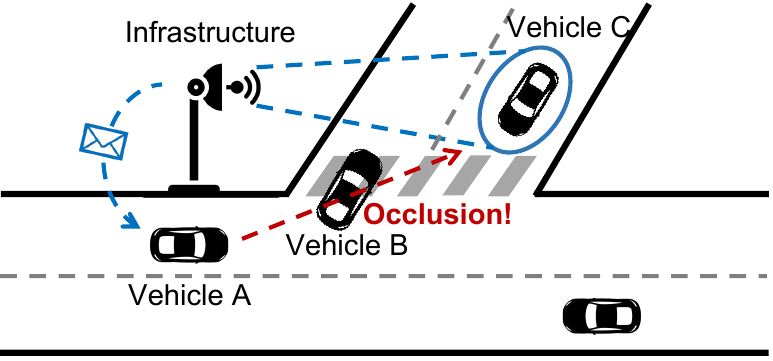}
    \caption{\textbf{Illustration of V2I cooperative perception.} From Vehicle A's perspective, Vehicle C is occluded by Vehicle B. By utilizing information sent by the roadside infrastructure, Vehicle A is enabled to detect Vehicle C. This capability enhances safety by providing a more comprehensive understanding of the surrounding environment.
    }
    \label{fig:enter-label}
\end{figure}

One of the primary challenges in V2I cooperative perception is achieving an optimal balance between perception accuracy and communication overhead. 
Transmitting the substantial volume of raw data captured by infrastructure sensors can lead to excessive communication overhead. 
Furthermore, the wireless network edge is often characterized by limited bandwidth and a highly dynamic communication channel \cite{shen2024large,shi2020communication}. 
Given the stringent latency requirements of autonomous driving, how to effectively compress the raw data for efficient transmission while maintaining the performance of cooperative perception is crucial.

In this paper, we propose TOCOM-V2I, a \emph{Task-Oriented COMmunication framework for V2I cooperative perception}.
The key idea is to reduce bandwidth consumption by transmitting only task-relevant information for the perception tasks, while filtering out irrelevant or redundant details \cite{shao2021learning}.
Specifically, a vehicle-side network and an infrastructure-side network independently extract features from their raw data.
Then, the infrastructure utilizes a spatial-aware feature selection module to identify the task-relevant information and adopts a hierarchical entropy model to exploit redundancy within the selected features for efficient transmission.
Finally, a scaled dot-product attention architecture fuses the features of both sides, and the vehicle perceives nearby objects based on fused feature maps.
Extensive experimental results show that the proposed TOCOM-V2I method outperforms previous works by achieving a better tradeoff between the perception performance and the communication overhead.
Our study demonstrates that the task-oriented design principle is promising to largely reduce the bandwidth consumption in V2I cooperative perception. 

\section{Related Work}

\subsection{V2I cooperative perception}

One key enabler of intelligent transportation systems is V2I cooperative perception, where information is shared from the roadside infrastructure to vehicles to extend the perception capabilities and enhance traffic safety.
Previous early fusion methods directly transmit raw observation data. While this preserves the detailed information, it consumes considerable bandwidth.
An alternative approach is late fusion, where both the infrastructure and the vehicle independently perceive the environment based on their raw data. 
The vehicle then fuses the perception results sent from the infrastructure.
While late fusion largely reduces the bandwidth consumption, it cannot effectively compensate for the missed detection by the vehicle, since the fine-grained details are lost during the independent processing.
To balance the communication overhead and the perception performance, intermediate fusion is a promising alternative, which extracts and transmits representative features from the raw data.
In particular, the authors of \cite{liu2020when2com,liu2020who2com} utilized handshake communication mechanisms that transmit the relevant features to match the requests.
Where2comm \cite{hu2022where2comm} introduces a spatial-confidence-aware communication strategy to share spatially sparse but perceptually critical information.
However, these approaches still face the problem of high communication overhead and latency.
As reported in \cite{hu2024pragmatic}, the intermediate fusion methods may lead to higher communication overhead than early fusion since they could have higher dimensions compared to the raw sensor data.
Besides, the cooperative perception systems involving multi-round communications introduce significant latency, making it difficult to meet the real-time requirements of cooperative perception in dynamic traffic scenarios.
Overcoming these challenges necessitates developing more efficient communication strategies that transmit and compress only task-relevant features for V2I cooperative perception.


\subsection{Task-oriented communications}

Driven by the interplay among artificial intelligence and communication networks \cite{wirelessllmshao}, recent investigation on task-oriented communications has gained significant attention \cite{shao2021learning,cai2023multi,li2023task,wen2023task,ShaoTWC}.
Traditional communication systems are $\emph{data-oriented}$.
They aim to guarantee the reliable transmission of every single bit, while oblivious to the message semantics or the downstream task.
For intelligent applications at the network edge, such communication systems will incur excessive delay if high-volume raw data are directly transmitted due to the constraints of limited communication resources.
The task-oriented design principle offers solutions to mitigate this issue.
Specifically, they discard information of little relevance to the task and prioritize the transmission of task-critical information.
By aligning the communication objectives with the task objectives, the task-oriented communication framework has the potential to achieve a better trade-off between communication efficiency and task performance.
There have been several recent studies taking advantage of this principle.
In particular, Xie et al. \cite{xie2023robust} developed a task-oriented communication scheme with digital modulation for image processing at the network edge.
Jankowski et al. \cite{jankowski2020deep} studied image retrieval over a wireless channel.
Shao et al. \cite{shao2023task} proposed a temporal entropy model to compress task-relevant features in edge video analytics.
In this work, we will explore the potential benefits of employing task-oriented communications within cooperative perception systems.

\begin{figure*}[t]
    \centering
    \includegraphics[width = 0.99\linewidth]{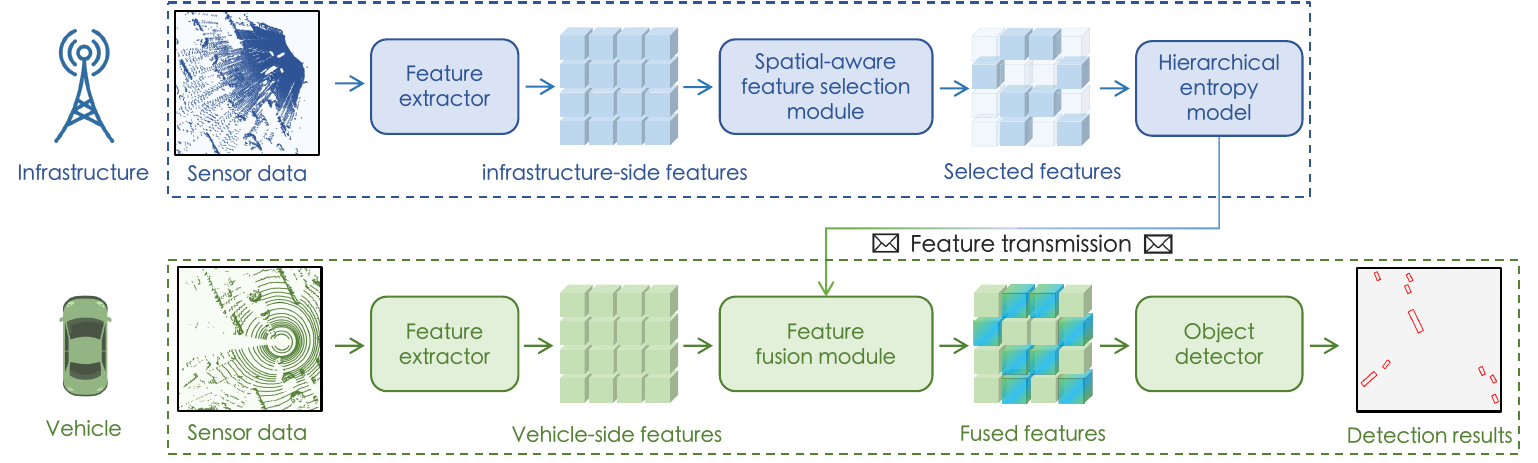}
    \caption{\textbf{The overall framework of TOCOM-V2I}.
    The spatial-aware feature selection module identifies the task-relevant information from the infrastructure-side features.
    The hierarchical entropy model exploits the redundancy within features for efficient compression and transmission of the selected features to the vehicle.}
    \label{fig:V2I-framework}
\end{figure*}

\section{System model and problem formulation}

The vehicle-infrastructure cooperative perception system consists of a vehicle and a roadside infrastructure.
The vehicle is equipped with onboard sensors such as depth cameras, LiDAR, and Radar to perceive its surrounding environment, while the roadside infrastructure, such as smart poles or traffic cameras, provides additional sensing capabilities from a different perspective. 
By leveraging the complementary perspectives, the cooperative perception system aims to fuse information from the vehicle and roadside infrastructure to enhance overall perception performance and robustness.
In this work, we consider the perception task of 3D object detection, and both the vehicle and the infrastructure capture LiDAR point cloud data from their viewpoints.

Formally, let $\bm{X}_{V}$ and $\bm{X}_{I}$ respectively represent the point cloud observations of the vehicle (V) and the infrastructure (I).
Denote $\bm{Y}$ as the perception supervision, and the objective of this system is to minimize a weighted sum of perception error and communication overhead:
\begin{equation}
\begin{aligned}
\min_{\Phi,\Psi}\ & D(\hat{\bm{Y}},\bm{Y}) + \beta * R(\hat{\bm{F}}_{I}), \\
& \text{with} \ \hat{\bm{Y}} := \Phi(\bm{X}_{V},\hat{\bm{F}}_{I}), \hat{\bm{F}}_{I} := \Psi(\bm{X}_{I}),
\label{equ:problem_formulation}
\end{aligned}
\end{equation}
where $\beta > 0$ is a coefficient.
The infrastructure leverages the model $\Psi$ to identify task-relevant features $\hat{\bm{F}}_{I}$ and send them to the vehicle.
The vehicle jointly uses the local data $\bm{X}_{V}$ and the received features $\hat{\bm{F}}_{I}$ for 3D object detection based on a cooperative perception network $\Phi(\cdot,\cdot)$.
$D(\cdot,\cdot)$ stands for the perception evaluation metric that measures the distance between the detection results $\hat{\bm{Y}}$ and the ground truth $\bm{Y}$.
$R(\hat{\bm{F}}_{I})$ represents the cost of transmitting the features $\hat{\bm{F}}_{I}$.
The main challenge of vehicle-infrastructure cooperative perception is the limited communication bandwidth at the wireless network edge, which makes it infeasible to transmit high-volume raw point cloud data from the infrastructure to the vehicle.
Therefore, we will adopt a task-oriented communication principle \cite{shao2021learning} that only transmits the features that are essential to the detection task, aiming to achieve a better tradeoff between the detection performance and the communication overhead.
The following section will describe our proposed method to address this challenge.

\section{Methodology}

The architecture of TOCOM-V2I is depicted in Fig. \ref{fig:V2I-framework}, which consists of feature extractors, a spatial-aware feature selection module, a hierarchical entropy model, a feature fusion module, and an object detector.
Feature extraction is the first stage, where raw data inputs are processed to produce a set of features. 
Then, the spatial-aware feature selection module evaluates the extracted features to determine their relevance for the perception task based on the spatial and contextual information.
Moreover, the selected features are quantized and encoded by entropy coding based on a hierarchical entropy model for efficient transmission.
After receiving the infrastructure-side information, the vehicle fuses the features of both sides for object detection.

\subsection{Feature extractor}

Consider that the vehicle and the infrastructure collect 3D point clouds $\bm{X}_{V}$ and $\bm{X}_{I}$ from their LiDAR sensors.
These raw data are rich in spatial information but redundant and unstructured.
The feature extractors aim to transform these high-dimensional raw data into compact and informative features, $\bm{F}_{V}$ and $\bm{F}_{I}$, for the detection task.
In this work, we leverage the Pillar Feature Network and the corresponding Backbone \cite{lang2019pointpillars} to extract visual features from point clouds.
We first project all the perceptual information to the same global coordinate system.
This unified representation avoids complex coordinate transformations and achieves better cooperation.
Then the raw point cloud is converted to a stacked pillar tensor and a pillar index tensor. 
The Pillar feature network uses the stacked pillars to learn a set of features.
These features are converted to a pseudo-image for further processing by a 2D convolutional backbone.
Such a backbone applies a series of convolutional layers to capture different levels of task-relevant information and follows a similar structure as \cite{lang2019pointpillars,zhou2018voxelnet}.
Specifically, it uses one network to produce features at increasingly small spatial resolution and leverages another one to concatenate the upsampled top-down features

\subsection{Spatial-aware feature selection module}

\begin{figure}[t]
    \centering
    \includegraphics[width = 0.99\linewidth]{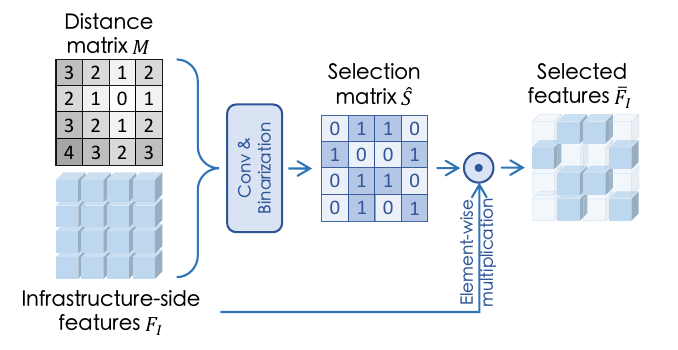}
    \caption{Spatial-aware feature selection module in TOCOM-V2I.}
    \label{fig:spatial-aware}
\end{figure}

As illustrated in Fig. \ref{fig:spatial-aware}, the spatial-aware feature selection module aims to identify features $\bar{\bm{F}}_{I}$ from infrastructure-side features $\bm{F}_{I}$ that are both perceptually significant and necessary for the vehicle.
In the context of object detection, the \mbox{areas} in $\bm{F}_{I}$ that contain objects are more task-relevant than the background areas.
During cooperation, these objects could help improve the perception quality by recovering occluded or missing objects in the vehicle's field of view.
Additionally, objects that are closer to the vehicle are more likely to be detected by the vehicle itself.
While these objects are perceptually important from an infrastructure perspective, there is no need to transmit these features.
Accordingly, the proposed module is spatial-aware when selecting the features, which utilize convolutional layers to generate a selection matrix $\bm{S}$ by taking both the extracted features $\bm{F}_{I}$ and a distance matrix $\bm{M}$ as inputs.
Specifically, each element $\bm{M}_{i, j}$ of $\bm{M}$ represents the distance between the location $(i,j)$ to the vehicle.
Each element $\bm{S}_{i}$ of $\bm{S}$ is in the range of $[0,1]$, where a higher value indicates that the feature at that location is more task-relevant. 
By quantizing each $\bm{S}_{i}$ to a binary value $\hat{\bm{S}}_{i} \in \{0,1\}$, we select the sparse feature map $\bar{\bm{F}}_{I}:= \hat{\bm{S}} \odot \bm{F}_{I} $, where $\odot$ stands for the element-wise multiplication.
Since the binarization operation is not differentiable, we use the {\color{black}straight-through estimator} \cite{bengio2013estimating_STE} to approximate the gradients during the training process.
Different from the previous studies \cite{hu2022where2comm,hu2024pragmatic} that use the classification head of the detector to find the important areas, our method enables joint optimization across all models in an end-to-end manner.
This has the potential to improve the performance of the cooperative perception systems.

\subsection{Hierarchical entropy model}

\begin{figure}[t]
    \centering
    \includegraphics[width = 0.99\linewidth]{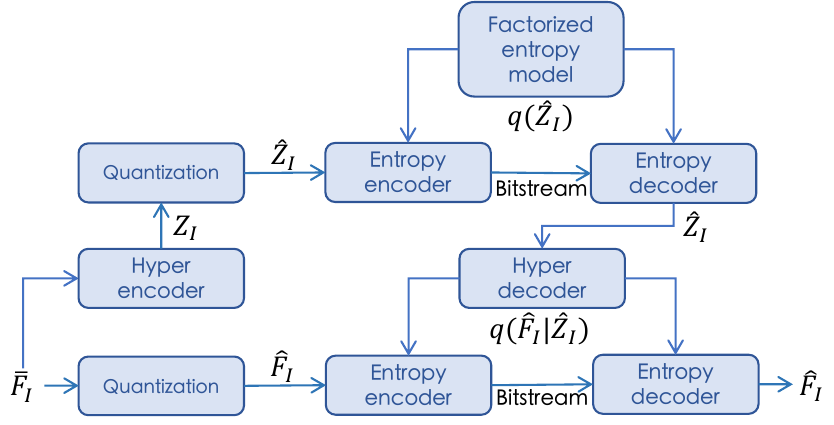}
    \caption{Hierarchical entropy model in TOCOM-V2I.}
    \label{fig:h_entropy_model}
\end{figure}

After identifying the task-relevant features $\bar{\bm{F}}_{I}$, the infrastructure quantizes these features to $\hat{\bm{F}}_{I}$.
To minimize the communication cost $R(\hat{\bm{F}}_{I})$, we apply entropy coding to $\hat{\bm{F}}_{I}$ for further compression.
As the entropy coding relies on a statistical model of the discrete variable, we employ learning-based methods to approximate the probability mass distribution $p(\hat{\bm{F}}_{I})$.
However, minimizing the loss function in (\ref{equ:problem_formulation}) faces a problem where the quantization operation produces zero gradients almost everywhere.
To allow optimization via gradient descent type algorithms in the presence of quantization, we follow the noise-based relaxation in \cite{balle2017end-to-end}.
The main idea is substituting additive uniform noise for the quantizer during the training process while switching back to actual quantization in the test phase.
In addition, we employ a hierarchical entropy model \cite{balle2018variational1} to estimate $p(\hat{\bm{F}}_{I})$.
This model incorporates a hyperprior to exploit dependencies in the feature space.
As shown in Fig. \ref{fig:h_entropy_model}, the hyper encoder transforms the features $\bar{\bm{F}}_{I}$ to hyper-latent representations $\bm{Z}_{I}$.
Subsequently, $\bm{Z}_{I}$ is quantized to $\hat{\bm{Z}}_{I}$, which is then compressed via entropy coding based on a fully factorized entropy model \cite{balle2017end-to-end}.
Such quantized representations serve as side information that allows the infrastructure to compress the features $\hat{\bm{F}}_{I}$ by a conditional entropy model $q(\hat{\bm{F}}_{I}|\hat{\bm{Z}}_{I})$.
Specifically, we model the distribution of $\hat{\bm{F}}_{I}$ as a Gaussian convolved with a unit uniform distribution, where the hyper decoder transforms $\hat{\bm{Z}}_{I}$ to the Gaussian parameters.

\subsection{Feature fusion and object detection}

The feature fusion module targets to augment the features $\bm{F}_{V}$ of the vehicle by aggregating the infrastructure-side features $\hat{\bm{F}}_{I}$.
To achieve this, we employ a scaled dot-product attention architecture to fuse the features at each spatial location, outputting the fused features $\bm{F}_{\text{fused}}$.
As the perceptual information is projected to the same global coordinate system, Such a location-wise fusion mechanism allows the vehicle to adaptively integrate information from the infrastructure-side features based on their relevance to the corresponding vehicle-side features.
This approach effectively exploits the spatial context while reducing computational complexity compared to global attention mechanisms.

Finally, the vehicle utilizes an object detector to output the detection results $\hat{\bm{Y}}$ based on the features $\bm{F}_{\text{fused}}$.
In particular, the detection function comprises two distinct branches. Each branch utilizes $1\times1$ convolutional layers for specific purposes: one for classifying the foreground-background categories, and the other for determining the bounding boxes.
Each location of $\hat{\bm{Y}}$ represents a rotated box, denoting the class confidence, position, size, and angle.

\section{Performance Evaluation}

\subsection{Experimental setup}

\textbf{Dataset:}
We conduct experiments on DAIR-V2X \cite{yu2022dair}, which is a real-world dataset for cooperative 3D Object Detection.
The cooperative frames are captured by a vehicle and an infrastructure-side unit.
Following \cite{hu2022where2comm}, we utilize additional labels in the 360-degree detection range and represent the field of view into a bird's-eye-view (BEV) map.
The perception range is set to $x \in[-102.4\mathrm{m}, 102.4\mathrm{m}]$ and $y \in[-38.4\mathrm{m}, 38.4\mathrm{m}]$.
The BEV feature map has the resolution of $0.4 \mathrm{~m} / \mathrm{pixel}$ in length and width.

\noindent \textbf{Baselines:}
We compare the proposed method against several baselines for cooperative object detection.
\textbf{Where2comm} \cite{hu2022where2comm} introduces spatial confidence maps to reduce bandwidth consumption and proposes confidence-aware message fusion to aggregate features to improve perception performance. 
\textbf{V2X-ViT} \cite{xu2022v2x} uses $1\times1$ convolutions to compress the feature maps along the channel dimension for efficient transmission and employs V2X-Transformer for information fusion.
To maintain fair comparisons, the communication round between the infrastructure and the vehicle is set to one.
In addition, we present \textbf{Early Fusion} as a communication cost upper bound and a performance upper bound, where the infrastructure directly sends the point clouds to the vehicle for cooperative perception.
Such a data-oriented communication scheme needs to reconstruct the raw data at the receiver, consuming a significant amount of bandwidth resources.

\noindent \textbf{Metrics:}
We mainly investigate the rate-performance tradeoff.
The rate refers to the data communication cost between an infrastructure and a vehicle.
For the 3D object detection task, we report average precision (AP) as the performance indicator, with intersection over union (IoU) thresholds of 0.5 and 0.3. 
This metric accounts for the normalized missed detections and false detections.

\noindent \textbf{Implementation details:}
All the methods are trained with a batch size of 2 on NVIDIA GeForce RTX 3090 GPUs. 
The Adam optimizer is selected for model training with an initial learning rate of 0.002 and weight decay of $10^{-4}$.
We improve the communication efficiency in cooperative perception by reducing the size of the feature maps and lowering the quantization resolution.
In our method, we set the hyperparameter $\beta$ in the range of $[10^{-4},10^{-2}]$.

\begin{figure}[!t]
     \centering
     \begin{subfigure}[t]{0.99\linewidth}
         \centering
         \includegraphics[width=\linewidth]{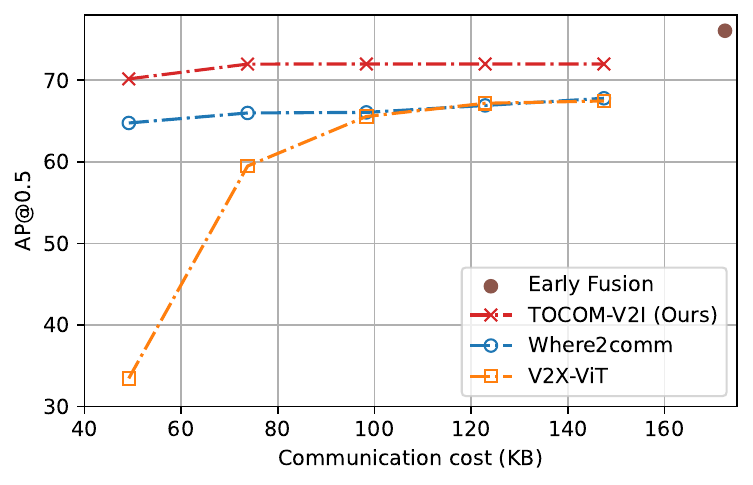}
         \caption{AP@0.5}
     \end{subfigure}
     \hfill
     \begin{subfigure}[t]{0.99\linewidth}
         \centering
         \includegraphics[width=\linewidth]{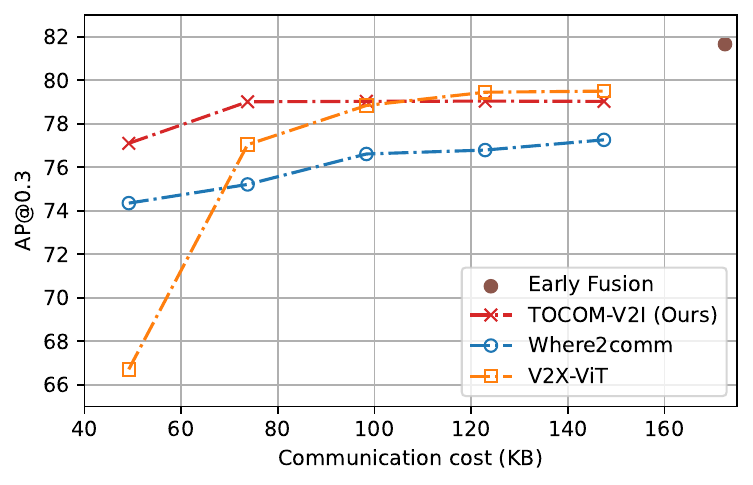}
         \caption{AP@0.3}
     \end{subfigure}
\caption{Performance-bandwidth tradeoff with (a) AP@0.5 and (b) AP@0.3.}
\label{fig:exp_results}
\end{figure}

\subsection{Experimental results}

Fig. \ref{fig:exp_results} illustrates the tradeoff between the detection performance and the communication overhead of the proposed TOCOM-V2I method and several baselines.
As anticipated, Early Fusion achieves the highest detection performance, given it has access to the raw 3D data of both sides.
However, its communication cost is prohibitively high, rendering it impractical for real-world applications where bandwidth is limited. 
Our method demonstrates a superior balance, achieving a better perception-communication tradeoff than all baselines. 
Specifically, TOCOM-V2I consistently gets a much higher AP@0.5 than other baselines for the same communication cost.
By setting the IoU threshold to 0.3, our method maintains comparable detection performance as the baselines at higher communication cost while achieving superior performance at lower communication cost.
This demonstrates the effectiveness of the proposed spatial-aware feature selection module and the hierarchical entropy model for essential information transmission.

\section{Conclusions}

This study verified the effectiveness of task-oriented communication for cooperative perception, which serves for 3D object detection rather than for raw data reconstruction.
Our proposed TOCOM-V2I enables a communication-efficient cooperative perception framework.
The core idea is to select the task-relevant information from the raw data based on a spatial-aware feature selection module.
Then the infrastructure adopts a hierarchical entropy model to exploit redundancy within the selected features for efficient transmission.
Finally, the vehicle leverages both the received perceptually critical features and its own observations to detect objects.
Extensive experimental results evidence that the proposed TOCOM-V2I achieves a better tradeoff between the detection performance and the communication overhead.

\bibliographystyle{IEEEbib}
\bibliography{refs}

\end{document}